\documentclass[prl,twocolumn,showpacs,amsmath,amssymb]{revtex4}

\usepackage{graphicx}
\usepackage{dcolumn}
\usepackage{bm}

\begin{document}


\title {Ultrafast Photoinduced Softening in a III-V Ferromagnetic Semiconductor\\
for Non-thermal Magneto-Optical Recording}
\author{G. A. Khodaparast, J. Wang, and J. Kono}
 \thanks{Author to whom correspondence should be addressed}
 \email{kono@rice.edu}
\affiliation{Department of Electrical and Computer Engineering,
Rice Quantum Institute, and Center for Nanoscale Science and Technology,
Rice University, Houston, Texas 77005}

\author{A. Oiwa}
\thanks{Present address: PRESTO, Japan Science and Technology
Corporation, 4-1-8 Honcho, Kawaguchi, 332-0012, Japan.}
\author{H. Munekata}
\affiliation{Imaging Science and Engineering Laboratory, Tokyo
Institute of Technology, Yokohama, Kanagawa 226-8503, Japan}

\date{\today}

\begin{abstract}
Through time-resolved two-color magneto-optical Kerr spectroscopy
we have demonstrated that photogenerated transient carriers decrease the
coercivity of ferromagnetic InMnAs at low temperatures.
This transient ``softening'' persists only during the carrier lifetime
($\sim$ 2 ps) and returns to its original value as soon as
the carriers recombine to disappear.  We discuss the origin of this
unusual phenomenon in terms of carrier-enhanced ferromagnetic exchange
interactions between Mn ions and propose an entirely nonthermal
scheme for magnetization reversal.
\end{abstract}
 
\maketitle


Mn-doped III-V semiconductors offer a variety of new scientific and technological
opportunities through their carrier-induced ferromagnetism
\cite{Munekataetal89PRL,Ohnoetal92PRL,Munekataetal93APL,
Ohnoetal96APL,Ohno98Science}.
Their magnetic characteristics are sensitive
functions of carrier density, and hence, carrier-density-modulation
can lead to control of ferromagnetism.
In fact, voltage-tuned \cite{Ohnoetal00Nature,Chibaetal03Science} and photoinduced
\cite{Koshiharaetal97PRL,Oiwaetal01APL,Oiwaetal02PRL} alteration
of ferromagnetism have been demonstrated recently.
In parallel to these pioneering studies on ferromagnetic semiconductors
are efforts to manipulate ferromagnetic metals (or itinerant ferromagnets).
A number of recent time-domain optical studies of
itinerant ferromagnets have revealed novel phenomena,
including ultrafast demagnetization \cite{Zhangetal02Book}.
However, there have been few time-dependent optical studies
of (III,Mn)V ferromagnets.

Here, we report on a demonstration of ultrafast photoinduced softening (UPS)
(i.e., a transient photoinduced decrease of coercivity) in ferromagnetic
InMnAs/GaSb heterostructures.  Pumping (III,Mn)V systems with ultrashort
laser pulses has a multitude of effects that are not present in the case of
ferromagnetic metals.  Since laser pulses can create a transient
distribution of delocalized carriers, one can anticipate significant
modifications in the exchange interaction between localized Mn spins and
delocalized carrier spins.
This should be contrasted to the metal case where the main effect of the
laser pulses is heating (either electronic or lattice heating).
We observed that photogenerated transient carriers significantly
{\em decrease} the coercivity ($H_c$).
We attribute this phenomenon to the carrier-enhanced
Mn-Mn exchange interaction and propose a new, extremely fast scheme
for recording
information on a magneto-optical disk entirely non-thermally.
\begin{figure}
\includegraphics [scale=0.45] {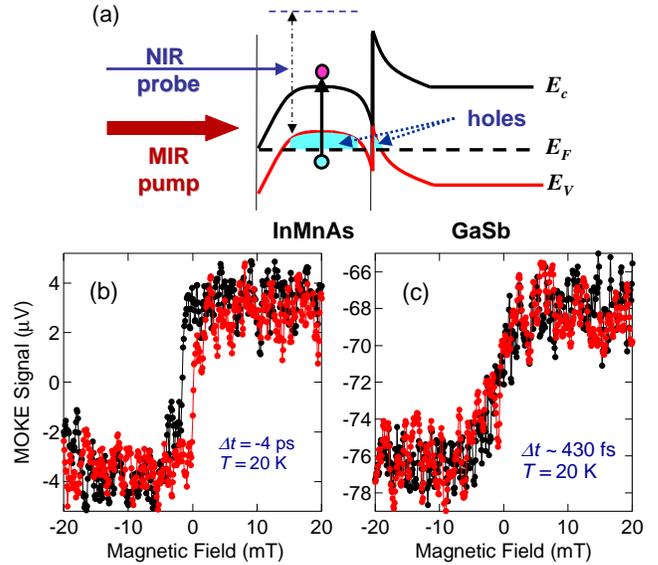}
\caption{(a) Band diagram of the InMnAs/GaSb sample.
We selectively create transient
carriers only within the magnetic layer using a mid-infrared pump.
A weaker near-infrared
beam, with photon energy far above the quasi Fermi energy of optical
excitation, probes the time-dependant ferromagnetism via Kerr rotation.
(b) and (c):
Ultrafast photoinduced softening.  MOKE signal vs.
magnetic field for two different time delays: (b)
$-$4 ps and (c) 430 fs.
In (c), the loop is nearly
destroyed in the horizontal direction, i.e., coercivity is almost zero.}
\label{scheme}
\end{figure}

Figure \ref{scheme}
shows a schematic band diagram of the InMnAs/GaSb sample studied.
Surface pinning of the Fermi energy and the type-II broken-gap alignment
between InMnAs and GaSb produces large band bending.
Also shown in
Fig. \ref{scheme} is the selective pumping scheme we used.  At the pump
wavelength (2 $\mu$m), the photon energy (0.62 eV) was smaller than
the band gaps of GaSb (0.812 eV) and GaAs (1.519 eV) but larger than that
of InMnAs ($\sim$ 0.42 eV),
so the pump created carriers only in the InMnAs layer (Fig.~\ref{scheme}).
Under our pumping conditions, the maximum density of photocreated carriers
are estimated to be comparable to or larger than the background carrier
density ($\sim$ 10$^{19}$ cm$^{-3}$), and, hence, significant modifications
in exchange interactions can be expected.

We performed two-color time-resolved magneto-optical Kerr effect (MOKE)
spectroscopy experiments using femtosecond ($\sim$ 150 fs) pulses of
midinfrared (MIR) and near-infrared (NIR) radiation.
Details of the setup were described previously \cite{Wangetal03JSC}.
The source of intense MIR pulses was an optical parametric amplifier
(OPA) pumped by a regenerative amplifier
(Model CPA-2010, Clark-MXR, Inc.).  The OPA was tunable
from 522 nm to 20 $\mu$m using different
mixing crystals.  We used a very small fraction ($\sim$ 10$^{-5}$) of
the CPA beam (775 nm) as a probe and the output beam from the OPA tuned to
2 $\mu$m as the pump.  The two beams were made collinear by a
non-polarizing beam splitter and then focused onto the sample mounted
in a magnet with optical windows.  We recorded the
intensity difference of the $s$- and $p$-components of the reflected NIR
beam as well as its total intensity as functions of time delay and
magnetic field.

The sample studied was an InMnAs/GaSb single
heterostructure with a Curie
temperature of 55 K, consisting of a 25 nm thick
In$_{0.91}$Mn$_{0.09}$As magnetic layer and an 820 nm thick GaSb buffer
layer grown on a semi-insulating GaAs (100) substrate.  Its room temperature
hole density and mobility were 1.1 $\times$ 10$^{19}$ cm$^{-3}$ and
323 cm$^2$/Vs, respectively.  The magnetization
easy axis was perpendicular to the epilayer due to the strain-induced
structural anisotropy caused by the lattice mismatch between InMnAs and GaSb.
This allowed us to observe ferromagnetic hysteresis loops in the polar
Kerr configuration.

Typical data showing the ultrafast photoinduced softening process are
presented in Figs.~\ref{scheme}(b) and \ref{scheme}(c).  Here, two magnetic-field
scans exhibiting ferromagnetic hysteresis loops at 20 K are shown,
for $-$4 ps and 430 fs time delays, respectively.  The pump beam was
circularly polarized.  The data at $-$4 ps delay [Fig.~\ref{scheme}(b)]
shows a hysteresis loop with a finite coercivity ($\sim$ 1 mT).
However, at 430 fs [Fig.~\ref{scheme}(c)],
the hysteresis loop is suppressed in the horizontal
direction, i.e., the coercivity is almost zero.  It is important to note
that the data shows almost
no change [or a slight {\em increase} (< 10\%), if any] in vertical size,
within the sensitivity of our setup.
This leads us to conclude that simple
lattice heating cannot be the reason for the above observations since
raising the lattice temperature should result in loop shrinkage both
horizontally and vertically.
\begin{figure}
\includegraphics [scale=0.57] {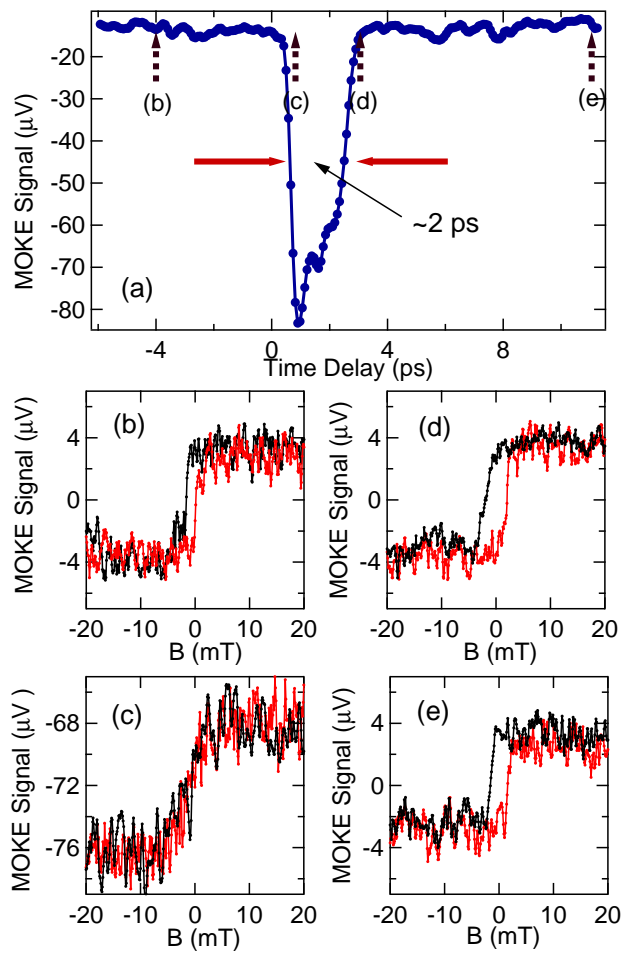}
\caption{Detailed MOKE dynamics.  (a) Time scan at $-$0.02 T and 20 K.
The 2-$\mu$m pump was circularly polarized ($\sigma^-$).  The probe
wavelength was 775 nm.  A $\sim$2-ps photoinduced MOKE
response is observed.  (b), (c), (d), (e): MOKE signals vs.
field at different time delays, corresponding to specific positions
shown in (a).  (b) $-$4 ps; (c) $\sim$ 500 fs; (d) $\sim$ 2 ps;
(e) 11 ps.
}
\label{evolution}
\end{figure}

Figure \ref{evolution} plots detailed photoinduced magnetization dynamics.
Figure \ref{evolution}(a) shows photoinduced MOKE signal versus time delay at a
magnetic field of $-$20 mT with a circularly polarized pump.  An ultrafast
photoinduced response is clearly observed.  Ferromagnetic loops are
plotted in Figs.~\ref{evolution}(b), \ref{evolution}(c), \ref{evolution}(d), and
\ref{evolution}(e), which correspond to the
fixed time delays indicated in Fig. \ref{evolution}(a).  Again, near
time zero [Fig.~\ref{evolution}(c)], the loop is collapsed horizontally.
It can be also seen that such softening lasts
only for a very short time, $\sim$ 2 ps.  As soon as the photoinduced MOKE
signal disappears [see Fig.~\ref{evolution}(a)], the loop returns with
the original $H_c$ recovered [see Figs.~\ref{evolution}(d) and
\ref{evolution}(e)].

First, let us discuss the origin of the ultrafast signal in Fig.~\ref{evolution}(a).
Note that none of the field scans in Figs.~\ref{scheme} and
\ref{evolution} are offset; i.e., vertical shifts of
the loops are a real effect.
However, we observed similar ultrafast photoinduced MOKE signals
even at temperatures far above the Curie temperature.
Therefore, we believe that the ultrafast vertical shifts
shown in Fig.~\ref{evolution}(a) are not due to any change related to
ferromagnetism.  We can also exclude nonlinear optical effects such as state
filling, band filling, and band gap renormalization as the cause
of the induced MOKE signal.  The two-color nature of the current
measurements allows us to decouple the photogenerated carriers from the
energy levels probed, i.e., the quasi Fermi level of the optically excited
carrier system is too low to affect the probe.
Finally, we found that the {\em sign} of the ultrafast MOKE signal depends on
the sense of circular polarization of the pump \cite{Wangetal03JSC},
i.e., $\sigma^+$ or $\sigma^-$.
These facts and considerations lead us to conclude that the
coherent spin polarization of the photogenerated
carriers is the origin of the ultrafast photoinduced
vertical shift of the ferromagnetic hysteresis loop.  Thus,
such time scans provide a direct measure of the spin lifetime of
photogenerated carriers.  In addition, we determined the charge lifetime of
the carriers to be $\sim$ 2 ps by standard pump-probe techniques.  The
short charge lifetime is probably due to anti-site defects introduced
during low temperature MBE growth \cite{Smithetal89APL},
but more systematic studies are necessary to understand the microscopic
mechanism of the ultrafast carrier recombination dynamics.

We believe that carrier-{\em enhanced} exchange coupling between Mn ions is
at the core of the observed ultrafast photoinduced softening.  The
phenomenon is essentially the same as what has been observed in the CW
work on the same systems \cite{Oiwaetal01APL,Oiwaetal01PE,Munekataetal02PE}
except for the very different time scales.  One possible scenario is in
terms of domain walls \cite{Oiwaetal01APL}.  Namely, creation of a
large population of transient holes breaks the original balance between
the exchange energy and the anisotropy energy (the latter was found to
be independent of carrier density \cite{Oiwaetal01APL}).
The dominance of the former results in an increase in domain wall
thickness and a decrease in domain wall energy.  This reduces the magnetic
field required to achieve magnetization reversal, i.e., coercivity is
decreased.  Another possible mechanism involves the magnetic rotation
of single-domain particles known as magnetic polarons.  In an assembly of
such particles, the interaction field among the particles is opposite to
the particle magnetization and
thus helps to reverse the particles.  This gives rise to a reduction of
the coercive field needed for magnetization reversal compared to that of a
single particle.
A large enhancement of the exchange coupling leads to the enlargement
of magnetic polarons and thus increases the single particle magnetization.
This increases
the interaction field and reduces the coercive force.
In either view, as soon as the transient
photogenerated carriers recombine to disappear, the original value of the
coercive field is recovered.  This is consistent with the notion that the extremely
short-lived photogenerated carriers are the cause of ultrafast photoinduced
softening.
\begin{figure}
\includegraphics [scale=0.48] {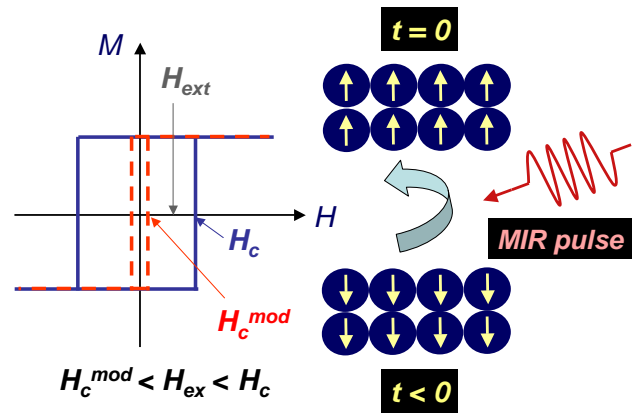}
\caption{Nonthermal magneto-optical recording by using ultrafast
softening in a (III,Mn)V semiconductor.
In a field ($H_{ext}$) smaller than the original coercivity ($H_c$) but
larger than the
expected photo-modified coercivity ($H_c^{mod}$), ultrafast creation of
transient carriers can
cause magnetization reversal.}
\label{device}
\end{figure}

Finally, we propose a new scheme for nonthermally recording information
onto a magneto-optical disk based on the observed ultrafast softening
(Fig.~\ref{device}).  If a storage medium made of a III-V ferromagnetic
semiconductor is prepared in a magnetic field that is smaller than the
original coercivity but larger than the expected photo-modified
coercivity, the ultrafast creation of transient carriers can soften
the recording material and, thus can cause ultrafast magnetization
reversal.  Our experiments above
showed that the softening process happens and persists only for less
than 2 ps, after which the coercivity recovers its original value.
Within such a time scale, any heat transfer process between spin and
lattice subsystems, which takes more than 1 ns, is out of the picture.
This proposed scheme, being nonthermal, possibly exceeds the fundamental
limit to the data writing rate set by the thermo-magnetic process used
in conventional magneto-optical recording (which is 30 MB/s in the
state-of-the-art technology) \cite{Kaneko00book}.

In summary, we have demonstrated ultrafast photoinduced softening, i.e.,
a transient coercivity decrease, in an InMnAs/GaSb ferromagnetic
semiconductor heterostructure.  We found surprisingly fast
switching due to ultrashort carrier lifetimes in
ferromagnetic semiconductor InMnAs, exceeding the fundamental time
limitation set by the thermo-magnetic process.  The extremely fast
decay and recovery of the observed ferromagnetic softening process
differentiates itself from any known ultrafast
processes in ferromagnetic systems.  We attribute the observed softening to
the carrier-enhanced exchange interaction between Mn ions and believe that
further systematic studies will provide detailed information
on the microscopic mechanism of carrier-mediated ferromagnetism in
these semiconductors.

This work was supported by DARPA MDA972-00-1-0034,
NSF DMR-0134058 (CAREER), NSF DMR-0325474 (ITR), and NSF INT-0221704.



\end{document}